\documentclass[10pt,conference]{IEEEtran}

\IEEEoverridecommandlockouts

\usepackage[switch,columnwise]{lineno} 
\usepackage{cite}
\usepackage{amsmath,amssymb,amsfonts}
\usepackage{algorithmic}
\usepackage{graphicx}
\usepackage{textcomp}
\usepackage{xcolor}
\usepackage{booktabs}
\usepackage{multirow}
\usepackage[obeyspaces]{url}
\usepackage{enumitem}
\usepackage{microtype}

\ifCLASSOPTIONcompsoc
    \usepackage[caption=false, font=normalsize, labelfont=sf, textfont=sf]{subfig}
\else
\usepackage[caption=false, font=footnotesize]{subfig}
\fi

\usepackage[disable]{todonotes}

\begin{document}
	
    \title{
        Industrial Control via Application Containers:\\
        Migrating from Bare-Metal to IAAS
		\thanks{This research has been partially funded by the Italian Ministry of Research and education under the program PRIN2015 and iCyPhy, under Siemens sponsorship.
		The authors wish to thank Pasquale Antonante for his help in the early phases of this project.}
	}

	\author{\IEEEauthorblockN{Florian Hofer,~\IEEEmembership{Student Member, IEEE}}
		\IEEEauthorblockA{\textit{Faculty of Computer Science} \\
			\textit{Free University of Bolzano-Bozen}\\
			Bolzano, Italy \\
			florian.hofer@stud-inf.unibz.it}
		\and
		\IEEEauthorblockN{Martin A. Sehr}
		\IEEEauthorblockA{\textit{Corporate Technology} \\
			\textit{Siemens Corporation}\\
			Berkeley, CA 94704, USA\\
			martin.sehr@siemens.com}
		\and
		\IEEEauthorblockN{Antonio Iannopollo,~\IEEEmembership{Member, IEEE}}
		\IEEEauthorblockA{\textit{EECS Department} \\
			\textit{University of California}\\
			Berkeley, CA 94720, USA\\
			antonio@berkeley.edu}
		\and
		\IEEEauthorblockN{Ines Ugalde}
		\IEEEauthorblockA{\textit{Corporate Technology} \\
			\textit{Siemens Corporation}\\
			Berkeley, CA 94704, USA\\
			ines.ugalde@siemens.com}
		\and
		\IEEEauthorblockN{Alberto Sangiovanni-Vincentelli,~\IEEEmembership{Fellow, IEEE}}
		\IEEEauthorblockA{\textit{EECS Department} \\
			\textit{University of California}\\
			Berkeley, CA 94720, USA\\
			alberto@berkeley.edu}
		\and
		\IEEEauthorblockN{Barbara Russo}
		\IEEEauthorblockA{\textit{Faculty of Computer Science} \\
			\textit{Free University of Bolzano-Bozen}\\
			Bolzano, Italy \\
			barbara.russo@unibz.it}
	}

	\maketitle

	\begin{abstract}
        We explore the challenges and opportunities of shifting industrial control software from dedicated hardware to bare-metal servers or cloud computing platforms using off the shelf technologies.
        In particular, we demonstrate that executing time-critical applications on cloud platforms is viable based on a series of dedicated latency tests targeting relevant real-time configurations.
	\end{abstract}


\section{Introduction}


Emerging technologies such as the Internet of Things and Cloud Computing are re-shaping the structure and the control of industrial processes in a radical way. 
These innovations allow the creation of highly flexible production systems, an essential component of 
the fourth industrial revolution.
Key enabling technologies such as distributed sensing, big-data analysis and cloud storage are taking the central stage in developing new industrial control systems.
Consequently, Edge Computing including cross-platform operation, third party software and mixed criticality applications are increasingly more important.
The computation requirements given the migration of functionality towards the "edge" imply  new system architectures~\cite{Telschigetal2018, Hofer2018}. 

The control of industrial processes, however, has not changed much over the last few decades, and there are reasons for it.
For instance, typical control applications found in industrial processes have to respond to changes in the physical world within predefined time limits.
Moving the execution of control tasks from devices physically co-located with the controlled process to cloud, egde or fog computing platforms requires dealing with 
delays that are difficult to predict.
Moreover, while dedicated control hardware and bare-metal solutions let the control design full authority over the environment in which its software will run, it is not straightforward to determine under what conditions the software can be executed on cloud computing platforms due to resource virtualization.
Yet, we believe that the principles of Industry 4.0 present a unique opportunity to explore complementing traditional automation components with a novel control architecture~\cite{Tascietal2018}.

We believe that modern virtualization techniques such as application containerization~\cite{Mogaetal2016,Tascietal2018,GoldschmidtHauck-Stattelmann2016} are essential for adequate utilization of cloud computing resources in industrial control systems.
The use of containerized control applications would yield the same advantages that traditional containerized microservices present: light and easily distributable control applications would be able,  for instance, to run on any system and, at the same time, be easily maintained and updated~\cite{Fazioetal2016}.

Beyond the migration capabilities and flexibility, containers simplify also the parallel execution of control software on devices such as PLCs and, to a lesser extent, on sensing and actuating field devices.
This results in increased reliability and robustness, while enabling further exploitation of self-* properties, including self-awareness and self-comparison~\cite{Lee2015}.
Time-machines (snapshots of the control software and/or of the machine state),
control redundancy (parallel operation of containers and/or of virtual server instances)~\cite{Lee2015}
and online system reconfiguration (reprogramming of machine's control algorithms and product specifications with no or reduced downtime)~\cite{TelschigKnapp2017}
are only a few of the Industry $4.0$ tools that will be made accessible.
Containers allow applications such as performance and distributed health monitoring~\cite{Wuetal2017,Terrissaetal2016} to run on a shared end node, and Digital-Twin~\cite{Schroederetal2016} to predict malfunction, maintenance intervals and tool lifespan.
Lastly, the application of modern virtualization techniques enables mixed criticality contexts which allow increased efficiency, reduction of the operational cost and decrease of production downtime~\cite{Royetal2016}. 

In this paper, we explore the feasibility of relocating real-time control applications, using off-the-shelf technology, from dedicated infrastructure and hardware onto a shared resource environment, both on a bare-metal host and in the cloud.
The contributions of this paper are: 
\begin{itemize}
    \item Survey of the state-of-the-art operating systems that support application containerization, which potentially allows the enforcement of real-time constraints. 

    \item Latency and determinism tests to identify parameters and configurations enabling the execution of control software on shared virtualized hosts.

    \item Comparison of virtualization instances and latency performances depending on configuration optimization.

	\item Evaluation of hard real-time task scheduling as application containers and demonstration of how, under specific conditions, the same tasks can be run in the cloud.
\end{itemize}

The rest of this paper is structured as follows. Section~\ref{sec:relwork} analyzes related work, while Section~\ref{sec:bgnd} introduces the background motivating our analysis. 
We then discuss the methodology and design of experiments, and review containerization frameworks in Sections~\ref{sec:method} and \ref{sec:contfram}.
We compare candidate host operating systems able to run the container engine in Section~\ref{sec:ostest}. Finally, we document the tests performed in Section~\ref{sec:virtsust}. 

\section{Related work}
\label{sec:relwork}

Containerizing control applications has been discussed in recent literature. 
Moga et al.~\cite{Mogaetal2016}, for instance, presented the concept of containerization of full control applications as a means to decouple the hardware and software life-cycles of an industrial automation system.
Due to the performance overhead in hardware virtualization, the authors state that OS-level virtualization is a suitable technique to cope with automation system timing demands.
They propose two approaches to migrate a control application into containers on top of a patched real-time Linux-based operating system: 
\begin{itemize}
    \item A given system is decomposed into subsystems, where a set of sub-units performs a localized computation, which then is actuated through a global decision maker. 
    \item Devices are defined as a set of processes, where each process is an isolated standalone solution with a shared communication stack.
    Based on this, systems are divided into specialized modules, allowing a granular development and update strategy. 
\end{itemize}
The authors demonstrate the feasibility of real-time applications in conjunction with containerization, even though they express concern on the maturity of the technical solution presented.

Goldschmidt and Hauk-Stattelmann in~\cite{GoldschmidtHauck-Stattelmann2016} perform benchmark tests on modularized industrial
Programmable Logic Controller (PLC) applications.
This analysis analyzes the impact of container-based virtualization on real-time constraints.
As there is no solution for legacy code migration of PLCs, the migration to application containers could extend a system's lifetime beyond the physical device's limits.
Even though tests showed worst-case latencies of the order of $15ms$ on Intel-based hosts, 
the authors argue that the container engines may be stripped down and optimized for real-time execution.
In a follow-up work, Goldschmidt et al.~\cite{Goldschmidtetal2018}, a possible multi-purpose architecture was described and tested in a real-world use case.
The results show worst case latencies in the range of $1ms$ for a Raspberry PI single-board computer, making the solution viable for cycle times in the range of $100ms$ to $1s$.
The authors state that topics such as memory overhead, containers' restricted access and problems due to technology immaturity are still to be investigated.

Tasci et al.~\cite{Tascietal2018} address architectural details not discussed in~\cite{GoldschmidtHauck-Stattelmann2016} and \cite{Goldschmidtetal2018}.
These additions include the definite run-time environment and how deterministic communication of containers and field devices may be achieved in a novel container-based architecture. 
They proposed a Linux-based solution as host operating system, including both the single kernel preemption-focused PREEMPT-RT patch and the co-kernel oriented Xenomai. 
With this patch, the approach exhibits better predictability, although it suffers from security concerns introduced by exposed system files required by Xenomai. For this reason, they suggested limiting its application for safety-critical code execution. 
They analyzed and discussed inter-process messaging in detail, focusing on the specific properties needed in real-time applications.
Finally, they implemented an orchestration run-time managing intra-container communication and showed that task times as low as $500\mu s$ are possible.

The three solutions discussed above share one common aspect: emph{they were based on bare-metal configurations}. 
These solutions were a first step for the re-allocation of an embedded control software onto a dedicated infrastructure.
They did consider real-time constraints, but were limited to execution on physical hardware.
The present trend of exploting the flexibility of resource sharing through cloud computing, 
makes it important investigating whether real-time control applications may also be ran on a shared virtualized infrastructure, and analyzing their capabilities and limitations.

In 2014, Garcia-Vallas et al.~\cite{Garcia-Vallsetal2014} analyzed challenges for predictable and deterministic cloud computing.
Even though their focus is on soft real-time applications, certain aspects and limits can be applied to any real-time systems.
Merging cloud computing with real-time requirements is a challenging task; the authors state that the guest OS has only limited access to physical hardware and thus suffers from unpredictability of non-hierarchical scheduling, and thick stack communications. 
While there exist real-time enabled hypervisors, such as the paravirtualized RT-Xen with direct access to hardware, the shared resources still suffer from latencies that may make real-time execution impossible.
Felter et al. in~\cite{Felteretal2015} focused on identifying the performance of instances based on hardware virtualization via Kernel-based Virtual Machines (KVMs) and container OS-virtualization using the cross-platform capable Docker. 
The benchmarks confirm that Docker results in equal or better performance than KVMs in almost all cases.
Arango et al.~\cite{Arangoetal2017} analyzed three containerization techniques for use in cloud computing. 
The paper compares Canonical's Linux Containers (LXC), Docker and Singularity, an engine developed by Lawrence Berkeley National Laboratory, to a bare-metal application. 
In many aspects, the Singularity containers performed better, sometimes even better than the bare-metal implementation, but this is largely due to the blended approach of the engine; Singularity is an incomplete virtualization solution since it grants access to I/O operations without context changes.

In summary, containerization techniques have been tested with early positive results in a variety of contexts.
However, the discussed approaches either focus on a low latency high performance cluster (HPC)~(\cite{Garcia-Vallsetal2014,Felteretal2015}) or on hard real-time applications running on bare-metal~(\cite{Tascietal2018, GoldschmidtHauck-Stattelmann2016, Goldschmidtetal2018}).
A combination of \textit{containers executed on cloud resources and strictly time-dependent control application containerization} has not yet been examined.
Such a configuration would require a kernel that supports and \textit{exceeds} soft real-time guarantees obtained by low latency kernel flavors in use on HPCs with only limited environmental control.
We assess the feasibility of this approach using off-the-shelf technology. 

\section{Problem Statement}
\label{sec:bgnd}

The goal of our experiments is to explore whether off-the-shelf technology can help to migrate hard real-time applications from a dedicated bare-metal infrastructure to a virtualized setup. 
To gain insight in opportunities and limits of this approach, a set of operating systems as well as system modifications are evaluated.
The resulting configurations are then tested to verify their suitability to run a real-time-capable containerization environment. 
If successful, this will allow a redistribution of applications onto a smaller amount of computing resources, thus saving a significant amount of resources and by doing so, reducing the cost of operating a given system in an important way.

The real-time software running on the system will be characterized by a set of real-time control tasks.
In the literature, three categories of real-time applications~\cite{Buttazzo2011} have been analyzed:
\begin{description}
    \item[Soft] Computation value decreases with deadline overshoot; 
    \item[Firm] Computation value obtained during an overshoot is zero; 
    \item[Hard] Missed deadlines may have catastrophic consequences.
\end{description}
If a deadline is missed, the outcome of a given set of control tasks is impacted, but an additional effect is the delay that may be caused to the execution of other tasks.
Consequently, when analyzing a certain overshoot, we must consider not only a single delayed task, but also its effects on the remaining execution schedule.

The given parameters of a periodic real-time application also determine its run-time boundaries.
The relation between execution parameters and deadline $d_i$ of a task $i$ is given by:
\begin{equation}
    f_i + r_i = c_i \leq d_i \leq p_i.
\end{equation}
where $r_i$ is the total run-time and $f_i$ the wake-up or firing time. 
The former expresses the actual used computation time including interruptions by higher priority tasks while the latter is the time spent between the period start and the actual execution start of a task, i.e. the latency.
The sum of run-time and firing delays defines the total time required to obtain the computation outcome, $c_i$. 
If the sum is higher than the relative deadline, the resulting misbehavior of the controlled system might have catastrophic consequences.
Hence, the sum should never exceed the deadline for the migration to be sustainable.
Measurement of the execution latency ($f_i$) inside virtual environments will allow us assessing the run-time boundaries of a known application. 
If the resulting latency is low enough, the spare computation time of an exclusive resource could be shared with other tasks.
An edge node running mixed critically applications can now share the resource burden and optimize resources based on priorities reducing the overall operational cost. 

The achievable amount of resource sharing depends on system configuration and operational noise from higher priority tasks such as interrupts, I/O delays and latency (increase of $c_i$).
Monitoring and reducing latency and computation time are therefore necessary but not sufficient to guarantee determinism in a shared context. 
However, a successful low latency evaluation will give a first result for the feasibility of application migration.
The impact of I/O and system latency will be explored in future work.

\section{Methodology and Design of Experiments}
\label{sec:method}
To asses the feasibility of the migration, we have to explore the running context and execute qualifying latency tests.
We first review state-of-the-art operating systems that can provide both (hard) real-time and container framework support. 
In addition to operating systems targeted for server infrastructures, we also evaluate some lightweight operating systems.
The selected OS must exploit the given resources properly, allowing the hardware to perform at its best, while not increasing the burden of operation.
Then we perform latency tests to verify the suitability of specific container-based virtualization solutions.
We are interested in understanding how these values change as we modify the environment. 
In the target configuration, a group of tasks will run on the same shared resource. 
Thus, observing the starting delays of a real-time task could clarify how the task might behave in different operational situations.
In addition, by applying computational and I/O stress to the shared resource running the latency test program, we can examine the effects of increased computational effort and I/O introduced latency on the real-time parameters of the applications; this will allow setting an upper limit on shared resource systems under large loads.

In general, we compare bare metal with virtualization approaches using hypervisors of Type 1 (native).
The former is expected to perform better in latency but worse in resource economy.
Virtualized instances, yet, are better in resource economy but this flexibility comes at the price of limited hardware control. 
Note that a Type 1 hypervisor is running directly on top of hardware, and thus it has better performance, while a Type 2 hypervisor runs on top of an underlying operating system.
The results of the latency tests show whether virtualization for hard real-time applications is sustainable and to which extent, reflecting expected maximum latencies.

We perform latency tests across three different phases, each consisting of a specific configuration and virtualization technique. 
In the first phase, we investigate performance variations on a Type 2 hypervisor. 
Because of the performance degradation induced by the Type 2 hypervisor, this setting represents the worst-case scenario.
Applying our latency tests to a real-time kernel will give insight on suitable configurations and/or environments for execution of real-time containers on Type 1 machines.
The latency tests during this phase consist of two isolation experiments, using tools like Linux control groups, and system configurations, such as task and interrupt affinity. 
The first experiment comprises latency tests that gradually isolate the virtualized guest from the host system.
In the second, we observe the effects on latencies if the same isolation is performed in the guest OS.
The two experiments will allow to analyze latency behavior and the impact of a system configuration change.
In the second phase, based on the previous results, we test and compare latency on three other systems: 
\begin{itemize}
    \item a bare-metal server, establishing the baseline for comparison of latencies and reflecting the \emph{status-quo} in current industrial control systems; 
    \item a Type 1 hypervisor controlled virtual generic instance;
    \item and a Type 1 hypervisor controlled virtual compute-optimized instance.
\end{itemize}
Finally, in the third phase, we evaluate latency within a container running on a Type 1 hypervisor hosted on a compute-optimized server.

\section{Container Frameworks Review}
\label{sec:contfram}
To isolate the different applications from each other, and due to better performance compared to traditional virtualization (see e.g.~\cite{Felteretal2015}), a containerized approach has been chosen.

The main choices available are 
\emph{LXC/LXD}\footnote{~LXC/LXD homepage: https://www.linuxcontainers.org/},
\emph{Docker}\footnote{~Docker homepage: https://www.docker.com}
and \emph{Balena}\footnote{~Balena project homepage: https://www.balena.io}.
LXC/LXD is the default Linux-based containerization engine. 
While LXC achieve containerization through a set of kernel-level primitives, 
the more recent LXD offers similar functionality with a more user-friendly approach.
Docker is an open source software written in Go, initially based on LXC technology, that has now moved to either directly use  \texttt{libcontainer}, a containerization library to access the dedicated kernel routines, or a variety of other isolation techniques.
It has a container catalog management, an open API and an easy-to-use CLI.
Balena is an open source stripped down version of Docker hosted by \textsc{resin.io} which allows minimizing the resource requirements. 
It maintains most of the configurations and parameters allowing a switch to Docker and its more comprehensive features at any time.
LXD focuses on (stateful) system containers, also called infrastructure containers, whereas Docker focuses on ephemeral, stateless, minimal containers. 
The former is better suited for long-running software based on clean distribution images.
The latter is indicated for instances that are not upgraded or re-configured but instead replaced entirely when a new version is available~\cite{Arangoetal2017}.
The two models are not mutually exclusive; for instance, LXD can be used to provide a full Linux system in a Docker instance.
For its flexibility and ease of use, we chose Balena as the containerization engine for our test setup. 

\section{Operating Systems Review}
\label{sec:ostest}
In this section, we review different candidate operating systems, in search of a generic, non-proprietary, real-time capable OS that is compatible with x86 and x86-64 architectures and has seamless support for containerization. 
Candidates are:
\begin{itemize}
	\item \textbf{resinOS:} Operating system by \textit{resin.io};
	\item \textbf{Ubuntu Core:}, Operating system for IoT devices;
	\item \textbf{Xenomai 3:}, Co-kernel extension for Linux-based OSs;
	\item \textbf{PREEMPT\_RT:}, Kernel patch for Linux-based systems.
\end{itemize}
Excluded in this study is the lightweight CoreOS, as it comes pre-configured with a LXC/LXD container engine, which does not conform with our criteria.
Similarly, The RTAI\footnote{~The Real-Time Application Interface for Linux, https://www.rtai.org} patch has been excluded from this analysis since it uses an approach very similar to Xenomai, 
while Xenomai has better OS support and allows user-level hard real-time tasks~\cite{xen02}.
.

\textbf{resinOS} is a Yocto\footnote{~Customized distros for embedded systems, https://www.yoctoproject.org/} Project-based operating system designed by ``resin.io'' to run containerized applications on small systems. It has Balena pre-installed and features a cloud-based interface to manage different hosts and containers.
Its image ships without package manager and building tools, requiring manual patching and increasing maintenance effort. 
Furthermore, given that the image has been tuned for operation on small devices, operation on server infrastructure may not utilize maximum hardware capabilities and thus reduce the system performance.

\textbf{Ubuntu Core} is a project by Canonical intended to exploit the advantages of standard Ubuntu technology and run small and transactionally upgradeable applications on embedded devices.
The application, developed for end devices, is shipped as a software package called \textit{snap}\footnote{~Snaps are a recent standard feature of Ubuntu-based distributions, combining the advantages of a package manager and an application container.} and runs sand-boxed in a container-like environment while retaining the transparency of a traditional application. 
Unfortunately, similar to resinOS, Ubuntu Core is deliberately minimal and ships without package manager and building tools.
In addition, even though the snap feature is native to Ubuntu and a patched kernel might add the needed real-time properties, the binding with the host name and file-space prevents the use of application duplicates. 

\textbf{Xenomai} is an OS extension that can be used to support POSIX real-time calls in a standard Linux kernel. 
Xenomai integrates with Cobalt, a small real-time infrastructure that schedules time-critical activities independently of the main kernel logic. 
The interaction with an interrupt dispatcher (I-pipe) allows increased response time and performance and thus enables hard real-time execution. 
\cite{Tascietal2018} suggest that Xenomai, together with an interrupt pipeline patch, has the best performance among the real-time OSs considered here.
Through further extension with interface skins, non POSIX-compliant real-time software can also be configured to run on a Xenomai patched system~\cite{ScordinoLipari2008,xen02}. The base-OS can be lightweight but still fully featured, 
including strong Debian-based package support and commercial Canonical company support.
Given how interrupt handling is implemented using out-of-band interrupt handlers and thus immediate IRQ reception, Xenomai is an ideal candidate for hard real-time applications. 
The scheduling of real-time tasks is performed in the co-kernel and is thus easier to customize, highlighting a promising approach.
However, the patching is bound to kernel versions, meaning progress depends on patch development of I-Pipe.

\textbf{PREEMPT\_RT} is a real-time kernel project maintained by the Linux foundation that minimizes the amount of kernel code that is non-preemptible.
To that end, several substitutions and new mechanisms are implemented~\cite{ScordinoLipari2008}.
%
Preempt-RT requires a dedicated patch, recompiling and tuning of the kernel to support hard real-time properties. 
The additional slicing preformed to the kernel tasks allows faster preemption and a better control of the CPU scheduling.
Compared to Xenomai, the handling of interrupt flow can not be controlled to the same extent, thus limiting the achievable performance.
In particular, drivers must be tuned for real-time operation to have low process firing jitter.
Recently, there were some major improvements in the PREEMPT-RT performance\footnote{~Fayyad-Kazan et al.~\cite{Fayyad-Kazanetal2014} show a performance improvement of 35\% with kernel versions v3.6.6 versus v2.6.33.}.
Its mainline development tracks kernel distributions at a fast pace, indicating that the project is strongly followed and has a valuable community backup, visible through events such as the ``Real-Time Summit''~\cite{lfnd01}.

To proceed with our tests, and in conformance with the requirements stated above, we define the technical criteria to be fulfilled.
Firstly, the container host OS has to be easy to deploy, manage and configure. 
Ubuntu Core, for example, features automatic updates via snap technology, making deployment automatic and centralized. 
Unfortunately, Ubuntu Core does not ship with a package manager, making general software maintenance and configuration a snap bound process.
All the software needed on this OS must either be shipped via this new mechanism or added and compiled manually. 
resinOS presents even more limitations: the OS neither features a package manager, nor ships with any tool to build code. The image is rather minimalistic and focuses only on small devices. 
Ubuntu Server LTS, our choice for Xenomai and PREEMPT-RT, is a full-fledged operating system. 
It features snap and package manager. Any needed tool can be easily added. 
Since it has been developed for long-lasting service and support, we believe it to be the best choice considering this aspect.

Secondly, solutions are compared for performance, resilience and scalability. 
The Xenomai-patched version has the advantage of a managed interrupt pipeline, giving the real-time kernel control over its flow. 
This means that Xenomai is the better performing of the two LTS systems. 
Its two-kernel architecture enables better control over the real-time to non real-time balance in case of system overloads, making the solution more resilient.
At the same time, the Xenomai community recommends using no more than four active real-time cores.
Indeed, since the Cobalt core does not share locks, handling a higher number of cores presents significant performance degradation~\cite{xen02}.
If we focus on this issue, PREEMPT\_RT would be the best option given that real-time scheduling is done in a single kernel, keeping control over locks and non real-time applications. 
The downside of PREEMPT\_RT is that we do not have full control over interrupts and consequently need to pay particular attention when selecting hardware and devices.

Overall, the most promising approach appears to be the Ubuntu Server LTS with the PREEMPT\_RT patch. 
To have a baseline for comparison, all latency tests in the following section will also be performed on our second choice, Ubuntu Server LTS with Xenomai 3.

\section{Experiment setup and Results}
\label{sec:virtsust}

The hardware configuration considered for the first tests is a dual core, 4 thread, i7 Skylake (U) system. 
The device will be used for the offline tests, is constantly powered with CPU and power saving settings set to performance.
For the configurations for phase two we selected Aamazon web services (AWS) to host the cloud-based environments.
Recent virtual instances use a new hypervisor based on KVM, called \texttt{hvm},
which allows direct assignment and control of hardware and resources reducing the virtualization overhead.
The resulting virtual instances offer comparable HPC performance, but with greater flexibility and scalability~\cite{amazon01}.
We selected thus an AWS HVM Type 1 hypervisor based T3.xlarge generic and a C5.xlarge computation optimized instance respectively. 
Both AWS instances run on 4 virtual CPUs, shared resources and use a custom kernel set-up for Ubuntu.
The last configuration, the baseline, is performed on a bare metal server with two Xeon X5560 processors on 8 cores, 16 threads.

The latency tests will be run on Ubuntu LTS with kernel 4.9.51 using two testing tools. 
We use \textit{cyclictest}~\cite{rttests01} version 1.0 to measure the latencies of cyclic firing behavior of a real-time application, and \textit{stress}~\cite{stress01} to simulate load in the system.
The former is part of the \texttt{rt-tests} suite, and a frequently used tool for this purpose~\cite{GoldschmidtHauck-Stattelmann2016, Mogaetal2016}, permitting us to compare results, performance and test configurations.
The latter runs random computations to simulate resource load and targets a system load average of 100\%.
The particularity about stress is that, in addition to CPU load, we can use it to generate threads for memory allocations, I/O sync requests and disk stress on the target system.
A ``stressed'' system is a rather extreme overload case and defines therefore an upper limit.

Table~\ref{tab:latcomp} and Figure~\ref{fig:plot} show only part of the results, focusing on the tests that gave most differing outputs. 
The script executing all the tests, the installation scripts, and the all experiment data, technical details and results are available online~\cite{homep01}.

\subsection{Offline tests}
\label{sec:first}

\begin{table}[tb]
	\centering
	\caption{Latencies for the two main OSs and their configurations using a single test thread (values are in $\mu s$, $n > 1mln$)}
	\begin{tabular}{l l | c c c c}
		Test & OS & Min & Avg & $\sigma$& Max\\
		\toprule
		\multirow{3}*{Default} & Standard & 2 & 6115 & 1195 & 767757\\
		& Xenomai & 2 & 7368 & 1829 & 916010 \\
		& Preempt-RT & 2 & 8070 & 3247 & 1147255 \\\midrule%
		\multirow{3}*{W. stress} & Standard & 2 & 648 & 36 & 18206\\
		& Xenomai & 3 & 495 & 124 & 14146\\
		& Preempt-RT & 3 & 544 & 22 & 13215 \\\midrule%
		\multirow{3}*{Isolated (iso)} & Standard &  2 & 6019 & 998 & 833695\\
		& Xenomai & 2 & 6867 & 1744 & 939560 \\
		& Preempt-RT & 2 & 8265 & 1650 & 1026677\\\midrule%
		\multirow{3}*{\shortstack[l]{Isolated (iso)\\w. stress}} & Standard & 3 & 542 & 26 &  25759\\
		& Xenomai &  1 & 398 & 36 & 23517 \\
		& Preempt-RT & 2 & 414 & 13 & 15055 \\\midrule%
		\multirow{3}*{\shortstack[l]{Isolated \& nlb\\\& IRQ affinity\\(irq)}} & Standard & 2 & 5186 & 229 & 675079\\
		& Xenomai & 2 & 5248 & 1957 & 907277\\
		& Preempt-RT & 2 & 5302 & 227 & 891876 \\\midrule%
		\multirow{3}*{\shortstack[l]{Isolated \\\& nlb \& irq\\\& w. stress}} & Standard & 2 & 8987 & 1140 & 5108938\\
		& Xenomai & 1 & 10570 & 3181 & 739785 \\
		& Preempt-RT & 2 & 8202 & 867 & 68123 \\\midrule%
		\multirow{3}*{\shortstack[l]{Isolated\\--host iso}} & Standard &  3 & 2630 & 4173 & 891020\\
		& Xenomai &  3 & 1168 & 2832 & 744583 \\
		& Preempt-RT & 3 & 2448 & 3729 & 1059626\\\midrule%
		\multirow{3}*{\shortstack[l]{Isolated w.\\ stress\\--host iso}} & Standard & 3 & 814 & 26 & 599697\\
		& Xenomai & 4  & 320 & 103 & 17463 \\
		& Preempt-RT & 2 & 508 & 14 & 13638\\\midrule%
		\multirow{3}*{\shortstack[l]{Isolated\\--host iso nlb irq}} & Standard & 2 & 332 & 229 & 637471\\
		& Xenomai & 1 & 469 & 177 & 225183 \\
		& Preempt-RT & 3 & 675 & 146 & 575708 \\\midrule%
		\multirow{3}*{\shortstack[l]{Isolated\\w. stress\\--host iso nlb irq}}  & Standard & 2 & 3808 & 82 & 912092\\
		& Xenomai & 0 & 8242 & 3799 & 560539 \\
		& Preempt-RT & 2 & 4788 & 431 & 55459 \\\midrule%
		\multirow{3}*{\shortstack[l]{Isolated-nlb\\irq\\--host iso irq}} & Standard & 3 & 4902 & 5451 &  938179\\
		& Xenomai & 3 & 4490 & 5024 & 971686 \\
		& Preempt-RT & 3 & 3145 & 5564 & 1185556\\\midrule%
		\multirow{3}*{\shortstack[l]{Isolated-nlb\\irq with stress\\--host iso irq}} & Standard & 2 & 363 & 190 & 45312\\
		& Xenomai & 0 & 189 & 139 & 18306 \\
		& Preempt-RT & 3 & 177 & 117 & 400991\\
		\bottomrule
	\end{tabular}
	\label{tab:latcomp}
\end{table}

We executed a set of test to verify different configuration options. 
The tests considered all combinations for isolation, load balancing, IRQ affinity and system stress for both host and guest system relying on VirtualBox version 5.2.18.
Isolation is achieved through \textit{Control groups} (CGroups).
They are used in Linux for resource partitioning and CPU pinning (affinity selection), and by Balena \& Co. to isolate containers from hosts. 
To give the virtual machine proper computing power, we set up CPU isolation between guest OS and host processes and, if needed, real-time tasks from non real-time processes inside the guest.
Such configuration relates to an AWS instance, where the guest OS threads are passed directly onto the hardware through the hypervisor.
The physical core might then be scheduled to be shared with another virtual instance, but without the computation burden of a complete OS.
IRQ-affinity and load balancer are kernel settings that can be changed during run-time.
The deactivation of the latter avoids that the kernel scheduler moves not strictly resource bound tasks to new resources. 
The former permits us to change the CPU assignment of some interrupts, and avoids thus that such interrupts when triggered force the a real-time task to wait.
Although we followed the ``Linux Realtime''~\cite{lrt01} guidelines for most of the following system tests, in this comparison we do not evaluate full dynamic ticks. 
In fact, for this configuration to work, a CPU must not run more than one thread. 
However, to save resources we want to maintain more than one real-time application per CPU and thus discarded this option.

During this first series of latency tests, we observed that virtual instances may be pinned to a certain CPU-set, Table~\ref{tab:latcomp}.
The results of the isolation test confirm that tasks running in parallel on the VM Host influence the latency performance tests.
If we compare the results of host only isolation with host and guest isolation for ``Isolated'' and ``Isolated w. stress'' in Table~\ref{tab:latcomp}, we note that the performance without stress decreases to one third while the stressed version has only a slight performance loss.
Furthermore, in most cases Xenomai and Preempt-RT perform better when under stress, with the best performance levels observed when isolated with a load balancer.

It seems that OS virtualization increases idle noise, but techniques such as the energy saving ``dynamic ticks'' are not operative.
With a loop frequency of 1000Hz we maintain a constant high scheduler invocation rate, higher than the schedulers interrupt tick, not explainig the noise.
It can however be due to the activity of the load balancer. In addition, the noise appears to be increasing when IRQ-affinity is set on a non guest isolated system.
A reason  may be that each of the CPUs is not a physical core but only one of two threads of the hyper-threading instance of it. 
Moreover, with load balancing on guest without stress the low load allows the host scheduler to fill up the available CPU time with other threads and eventually to distribute them between CPUs.
This movement of tasks causes the CPU registers and L1-L2 caches to be flushed and refilled from the L3 cache. 
In addition, it is impossible to pin the single threads representing a virtual CPU to a physical CPU, even when shielded, so that the guest threads might be moved to other CPUs while interrupts ``steal'' an assigned CPU.
The latter characteristic seems to be isolated to the implementation of the standard VirtualBox hypervisor. 
All these factors introduce latencies that are visible in the average and peak values.
This may indicate room for further latency improvements by removing interrupts and overload also on the CPU siblings, influencing directly the maximum performance and average latency of the real-time tasks.
The guest under stress, yet, does not allow much free CPU-time on the physical CPUs instead. Thus, the host threads and all movable tasks are mostly run on the CPU not serving any of the virtual machine threads.
This ultimately prevents the scheduler from shifting threads to other CPUs, and thus avoids added latency.

The most interesting result is gathered when the guest is isolated and host is configured with isolation, no load balancer and IRQ affinity, Table~\ref{tab:latcomp} test 9 and 10.
In comparison with previous results, in this test the latency statistics invert, having better performance and less noise in the ``idle'' case.
As discussed before, the stressed test is more an extreme upper limit than an actual use case.
Thus, the decreasing latency performance depicts the expected behaviour allowing us to predict latency, and consequently determinism, with increasing system load, better.
In this setting, the PREEMPT-RT version performs best among the three kernel versions having the lowest maximum peaks under stress.
In addition, this configuration is actually closer to an isolated bare metal or a hypervisor Type 1 solution.
The presented configuration tries to prevent the host from interacting with the guests resources.
If we add the no-load balancer and IRQ affinity configuration for the guest system the results invert again.
Therefore, the results indicate that we should expect the best behavior with CPU isolation on bare metal solutions as well as hypervisor Type 1 products.

\subsection{Hardware comparison}
\begin{figure}[tb]
    \centering
    \includegraphics[width=\linewidth]{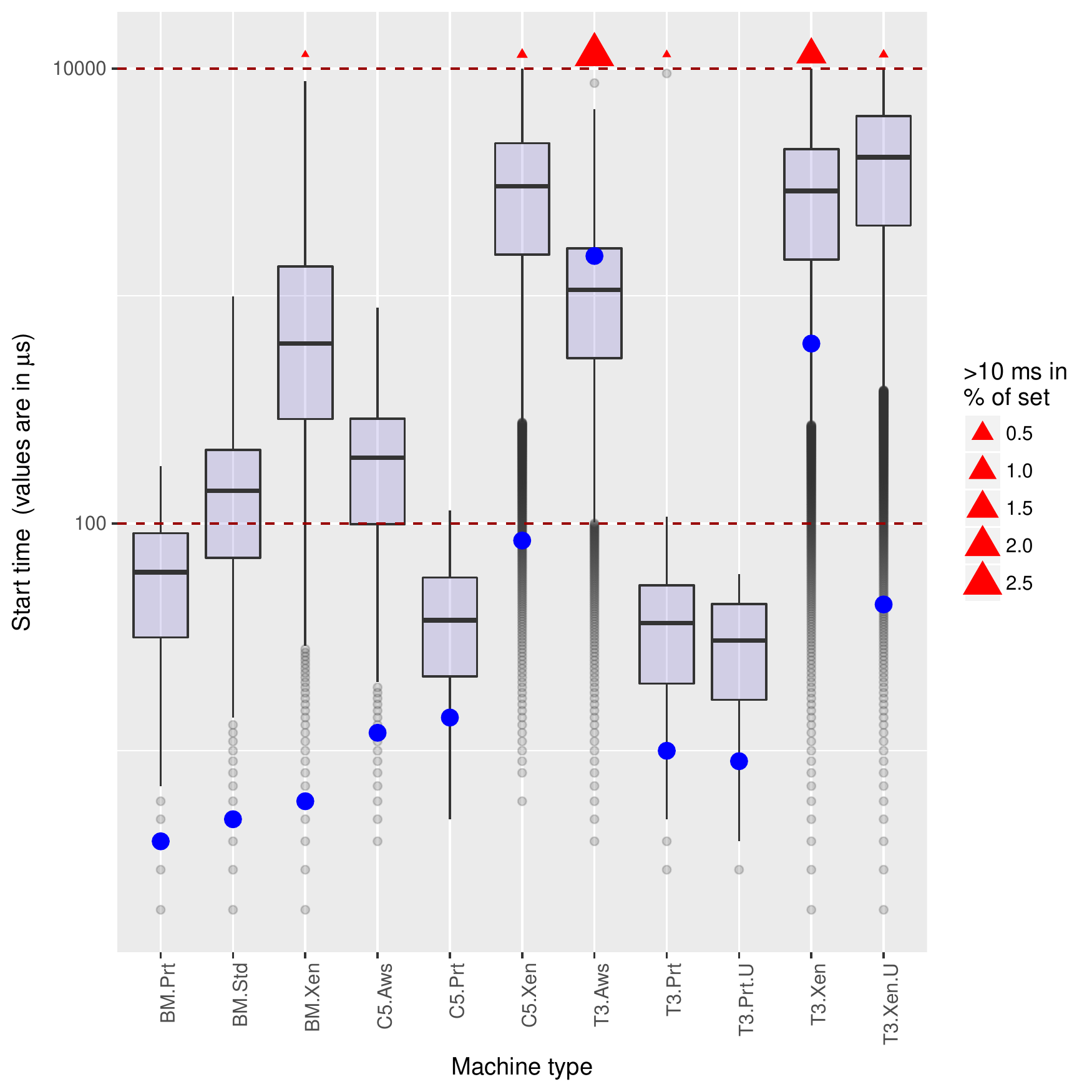}
    \caption{Boxplot of latencies, with averages and overshoot sizes - mean in blue}
    \label{fig:plot}
\end{figure}

In light of all tests performed here, we identified the following configuration as favorable for our purposes: 
Isolation, with load balancer.
Based on this configuration, we continue with a set of experiments on different hardware under stress.
The resulting latency values are summarized in Figure~\ref{fig:plot}.

A total of ten million loops over multiple hours have been executed for each configuration.
Different from our previous experiments, the number of \texttt{cyclictest} \textit{and} \texttt{stress} threads is given by the size of the isolation configuration, i.e. for each vCPU running real-time we have one thread of each kind.
Ideally, the maximum firing delay of the threads should stay below $\frac{1}{10}^{th}$ of the cycle time.
This upper limit has deliberately been set to allow minimal resource sharing for our sample application.
Therefore, we added two reference lines in the plot visualizing those thresholds, one at $10ms$ ($100ms$ cycle) and $100\mu s$ ($1ms$ cycle).

All obtained results have been gathered under \textit{stress} and should be considered as worst case scenarios.
Among all standard kernel configurations, the reference bare metal solution performs best in mean and spread. 
The same is true for Xenomai based versions. 
Note that the Xenomai configuration of the various systems always performs worst.
The two main reasons are: 
1) one configuration exposed more than the recommended 4 cores, causing a high amount of context switching overhead, 
and 2) hypervisor IRQ latency that removes the advantage of the dedicated interrupt pipeline of Xenomai compared to PREEMPT-RT and AWS/Standard kernels. 

On the other hand, if we consider the PREEMPT\_RT configurations in Figure~\ref{fig:plot}, the bare metal set-up still performs best in mean, but, this time, it has a higher spread than the C5 and T3 instances.
However, the average latency times are all in the range of 5 to $12\mu s$ and, in addition, each spread has the majority of values below our lower target of $100\mu s$.
Therefore, our first suggestion for a IAAS replacement can be a general T3 instance. 
With only 96 occurrences out of 10 million (0.00096\%) exceeding the upper limit, it can be an economic solution where strict determinism is not needed
or cycle times are higher than the peak value measured, $49ms$.  
However, in such cases the peaks should be managed properly via tools such as scheduling monitors and/or orchestrators.
If stricter limits are required, the C5 PREEMPT\_RT instance showed the lowest spread and peak ($114\mu s$) among the measured instances.
Interestingly, a test run with a PREEMPT\_RT T3-Unlimited enabled instance produced even better results.
The latter is a feature that allows CPU bursts for short high load peaks on T3 instances at additional price.
This economic variant can therefore also be used for applications requiring stricter determinism.
In conclusion, the results are promising and confirm the feasibility of migration to IAAS solutions.

\subsection{Test-run of a Balena container}

To use real-time control group capabilities inside Balena and Docker containers, an additional kernel configuration flag must be enabled. 
Unfortunately, the flag cannot be enabled for the PREEMPT\_RT version of the systems. 
Due to some temporary compatibility constraints, the simultaneous operation of both full kernel preemption and RT-CGroup has been disabled. 
Details can be found at~\cite{lfnd01}.

Finally, we successfully ran a test container with the real-time test program.
Latency tests executed as above confirms our expectations for the container: with average values of $7(\sigma 0$)/$11.44(\sigma 0.71)\mu s$  for stress/no stress respectively, and maximum peaks of $7655/11644\mu s$ in a default configuration for a C5 system, a migration to an IAAS virtualized instance is possible. 

\section{Conclusions and Future Work}
\label{sec:conclusion}
In this paper, we explored limits and feasibility of migrating real-time applications from bare metal servers to virtualized IAAS configurations.
We identified two promising operating systems, Xenomai and PREEMPT-RT patched Linux, and tested task latencies in a variety of configurations, with the system both in idle and stress scenarios. 
Although the tests aim only at computation latency, appropriate response times are a strict requirement even before considering I/O and system latency.
We observed that the performance of the latency tests on a Type 1 hypervisor is comparable to the performance on a bare metal solution.
Finally, we configured a real-time container engine and built and ran a containerized real-time application. 
Our experiments showed that the containers could be grouped and orchestrated using the CGroups feature of the Linux kernel to optimize resource utilization.

We showed that containerization introduces a novel paradigm for control applications. 
Previously isolated computation tasks, however, are now operating concurrently and interacting with each other, potentially influencing timing performance.
We concur with Goldschmidt et al.~\cite{Goldschmidtetal2018} that this new paradigm requires investigation on topics such as container security, restricted container access and intra-container data exchange. 
In the future, we are planning to consider orchestration tools that can schedule real-time containers based on pre-configured capacities.
The goal here is to maximize resource utilization without significantly impacting overall execution determinism, at first on computation only and finally considering I/O and system latency.
Moreover, we believe that latency and performance tests of recent releases of a patched Linux kernel should be further investigated.
Indeed, the latest releases introduce 
a new Earliest Deadline First scheduler for hard real-time applications.
However, the Linux kernel used by AWS, although optimized for their hardware, is not up to date and inline with the latest kernel releases.
Hence, this optimization might prevent the full utilization of new capabilities and create a configuration mismatch, affecting maximum achievable performance.
We believe that the proper configuration and tuning of the Linux kernel parameters may improve overall task determinism and is worthy of further investigation.

\bibliographystyle{IEEEtran}
\bibliography{IEEEabrv,bibliography}

\end{document}